\documentclass[journal,twoside,web]{ieeecolor}
\usepackage{jsen}
\usepackage{cite}
\usepackage{amsmath,amssymb,amsfonts}
\usepackage{algorithmic}
\usepackage{graphicx}
\usepackage{textcomp}
\usepackage{wrapfig}

\usepackage{array}
\usepackage{caption}
\usepackage{booktabs}
\usepackage{multirow}
\usepackage{graphicx}

\def\BibTeX{{\rm B\kern-.05em{\sc i\kern-.025em b}\kern-.08em
    T\kern-.1667em\lower.7ex\hbox{E}\kern-.125emX}}
\definecolor{abstractbg}{rgb}{0.89804,0.94510,0.83137}
\begin{document}
\title{Process Optimization and Deployment for Sensor-Based Human Activity Recognition Based on Deep Learning}
\author{Hanyu Liu, Ying Yu, Hang Xiao, Siyao Li, Xuze Li, Jiarui Li, Haotian Tang*
\thanks{*This is our corresponding author. Hanyu Liu, Hang Xiao, Siyao Li, Xuze Li, Jiarui Li, Haotian Tang are from Northeastern University,  Shenyang 110169, China (liu\_han-yu@foxmail.com, 20237348@stu.neu.edu.cn, 20237376@stu.neu.edu.cn, 20246350@stu.neu.edu.cn, 20246323@stu.neu.edu.cn, haotiantang@stumail.neu.edu.cn). Ying Yu is from Liaoning University,  Shenyang 110169, China (yas254540645@outlook.com).}}
\IEEEtitleabstractindextext{%
\fcolorbox{abstractbg}{abstractbg}{%
\begin{minipage}{\textwidth}%
\begin{wrapfigure}[12]{r}{3.1in}%
\includegraphics[width=3in,height=1.5in]{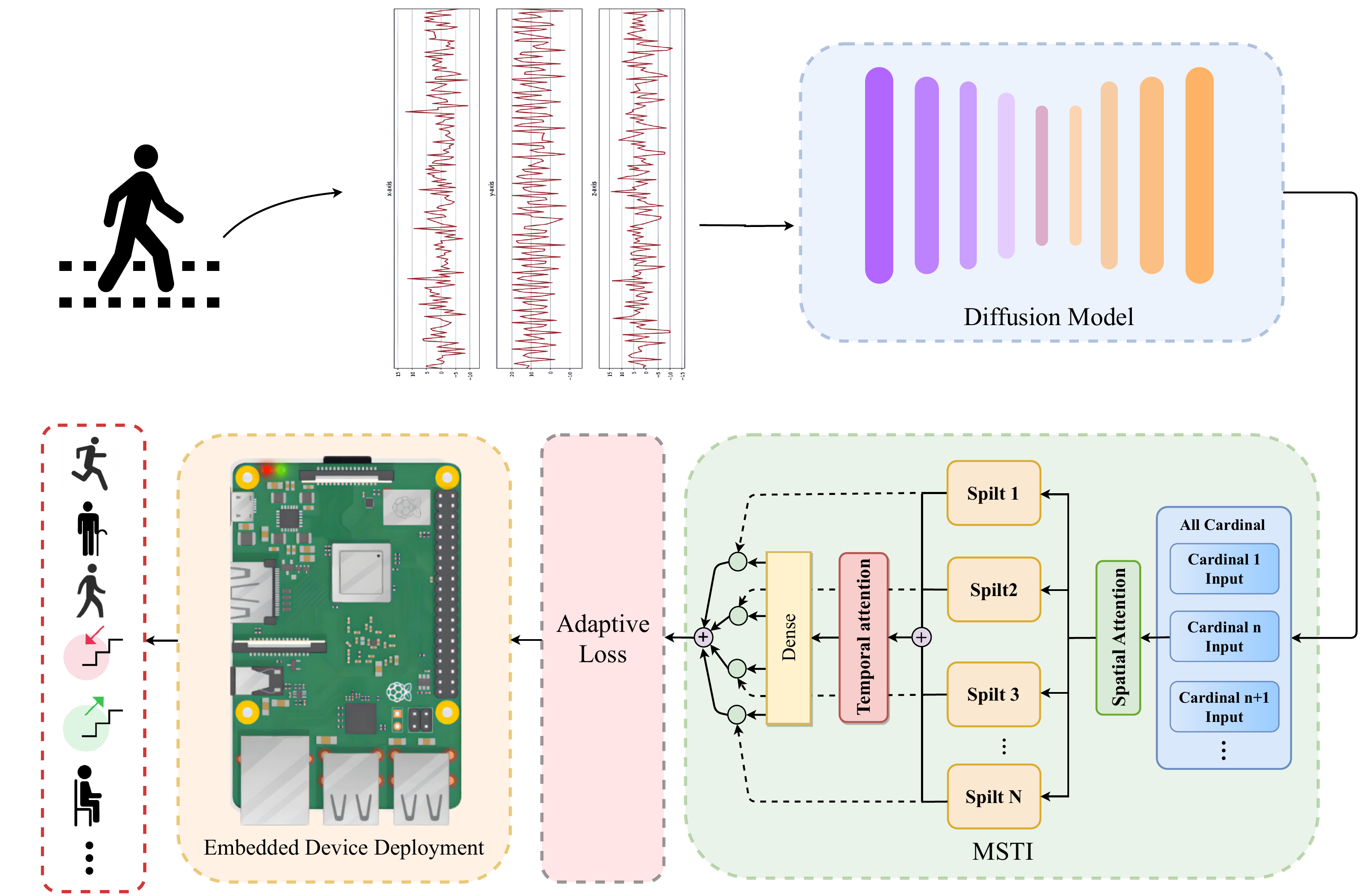}
 
\end{wrapfigure}%
\begin{abstract}
Sensor-based human activity recognition (HAR) is a key technology for many human-centered intelligent applications. However, this research is still in its infancy and faces many unresolved challenges. To address these, we propose a comprehensive optimization process approach centered on multi-attention interaction. We first utilize unsupervised statistical feature-guided diffusion models for highly adaptive data enhancement, and introduce a novel network architecture—Multi-branch Spatiotemporal Interaction Network (MSTI), which uses multi-branch features at different levels to effectively Sequential, which uses multi-branch features at different levels to effectively Sequential spatio-temporal interaction to enhance the ability to mine advanced latent features. In addition, we adopt a multi-loss function fusion strategy in the training phase to dynamically adjust the fusion weights between batches to optimize the training results. Finally, we also conducted actual deployment on embedded devices to extensively test the practical feasibility of the proposed method in existing work.  We conduct extensive testing on three public datasets, including ablation studies, comparisons of related work, and embedded deployments.
\end{abstract}

\end{minipage}}}

\maketitle

\section{Introduction}
\label{sec:introduction}
\IEEEPARstart{H}{uman} Activity Recognition (HAR) is an emerging field with broad application prospects, aiming to identify subjects' behaviors over time using motion information \cite{essa2023temporal}. Current HAR systems are predominantly video-based or sensor-based \cite{ijjina2017human}. Video-based systems recognize behaviors through captured images or videos, facing societal and technical challenges, such as privacy concerns, dependency on environmental lighting, resolution constraints, and the high cost and complexity associated with video processing algorithms. These challenges have significantly impeded their widespread adoption \cite{kumar2021human}. In contrast, inertial sensors, including accelerometers, gyroscopes, and magnetometers, have emerged as an ideal solution for HAR due to their privacy-preserving and convenient nature. These sensors are often embedded in wearable devices like smartwatches or gloves, offering compactness, precision, and cost-effectiveness, thus overcoming many limitations of environmental equipment \cite{yang2022activity}. With the proliferation of sensor devices, HAR has garnered considerable attention in the domain of pervasive computing. This field employs a variety of algorithms to interpret human activities, utilizing data collected from sensors attached to different parts of the body. This burgeoning research area has propelled the development of numerous context-aware applications, including healthcare, fitness monitoring, smart home technologies, and elderly fall detection \cite{yadav2022arfdnet}\cite{host2022overview}. 
Considerable research has been devoted to exploring HAR, initially adopting classical machine learning methods such as Decision Trees (DT) \cite{anguita2013public}, Support Vector Machines (SVM) \cite{vijayvargiya2021implementation}, Random Forests (RF) \cite{galvan2016analysis}, and Naive Bayes (NB) \cite{liu2013human}, favored for their low computational complexity and broad applicability. However, these methods are limited by the representational features they extract, constraining classification performance and suitability for smaller datasets. Various deep neural networks, such as Convolutional Neural Networks and Long Short-Term Memory networks, have become significant research topics in extensive HAR scenarios, exhibiting sustained superiority \cite{zeng2014convolutional,dang2020sensor}. Deep learning methods, compared to traditional machine learning techniques, can automatically extract deep feature representations from sensor signals, enhancing the accuracy of HAR \cite{wang2019deep}. 
Nevertheless, deep feature extraction for sensor-based HAR continues to pose serious challenges:
\begin{figure*}
    \centering
    \includegraphics[width=1\linewidth]{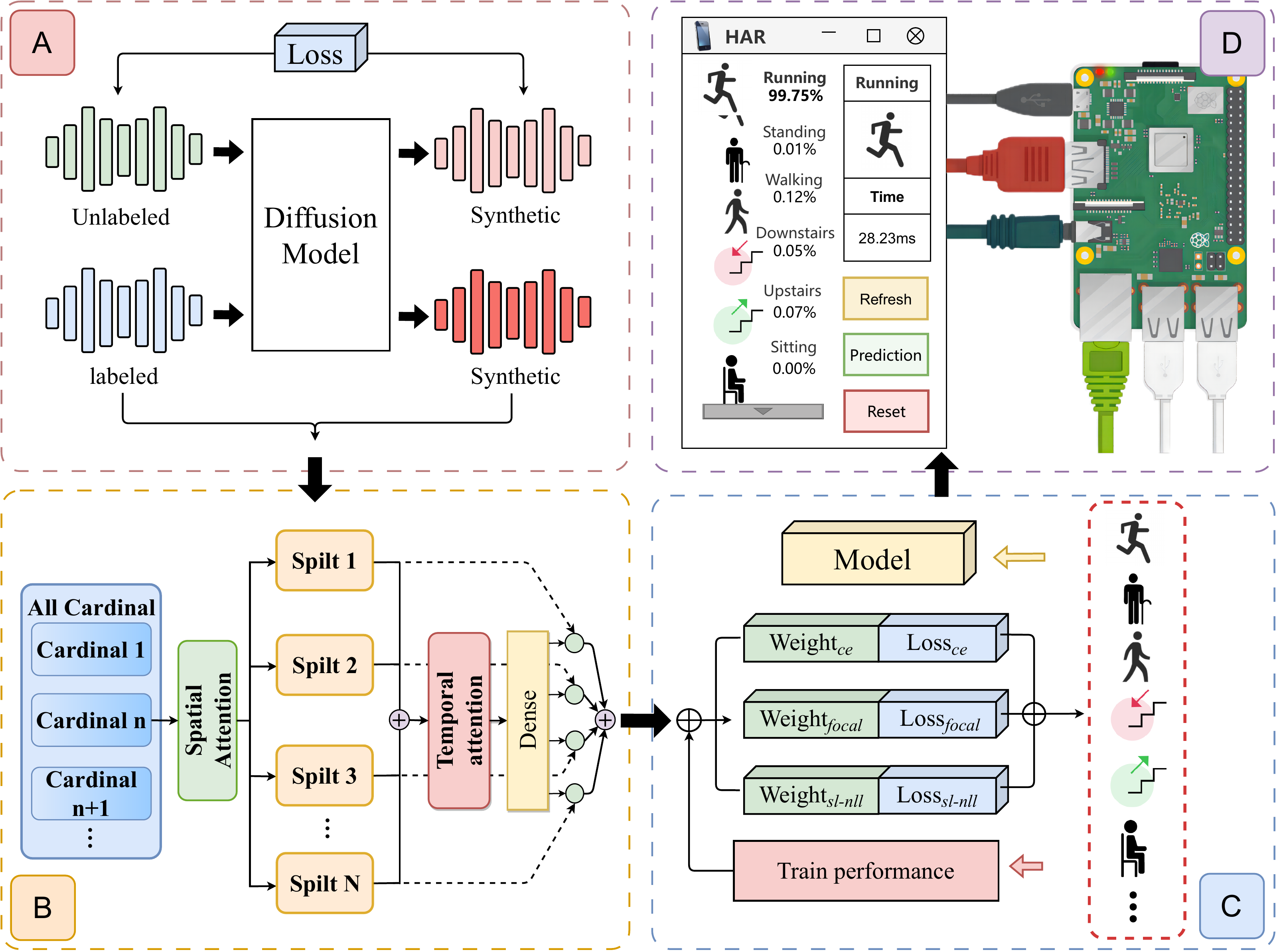}
    \caption{The total process of task. }
    \label{fig1:ALL}
\end{figure*}

\textbf{Deep Feature Extraction:} The diversity of human activities means that the same behavior may exhibit different feature patterns across various environments and contexts. For instance, the action of jumping to hit a ball in badminton is similar to that in volleyball, yet they are distinct activities. This often leads to reduced accuracy, longer development and training costs, and decreased robustness of models when applied \cite{liu2023mag}. 

\textbf{Class Imbalance:}Class Imbalance is inherent in label data, often presenting a long-tailed distribution \cite{cao2020heteroskedastic}. In the real world, some human activities occur more frequently than others, leading to class imbalance issues within datasets. This imbalance can result in poor recognition performance for minority classes, as they may be overlooked due to their smaller representation, causing the model to generalize poorly on the training set \cite{ahn2023cuda}. 

The remainder of this paper is organized as follows: Section 2 reviews some related methods in the field relevant to our work. Section 3 outlines the current workflow and discusses the implementation details of our method. Section 4 presents the experimental details. Finally, Section 5 exhibits the performance of the model in experiments and discusses the current research findings. 

\section{Related Work}
\subsection{Deep Feature Extraction}
The diversity, multimodality, high dimensionality, variability, and dynamism of human activities make deep feature extraction crucial for the accurate classification and recognition of these activities. Zhang et al. \cite{zhang2023attention} applied residual connections in HAR, combining spatial features extracted by 1D-CNN with bidirectional long short-term memory (BLSTM) through residual links, thus enhancing the model's capability to comprehend complex temporal patterns. The ResNet network has also been adapted as an underlying network for HAR. Ronald et al. \cite{ronald2021isplinception} improved and applied the ResNet network to HAR tasks, outperforming other deep learning architectures previously proposed on four public datasets. However, traditional ResNet models face issues such as poor inter-channel correlation, large parameter count, and insufficient feature reuse. To address these issues, Zhang et al. \cite{zhang2022resnest} proposed the ResNeSt network, which quickly gained widespread application upon its introduction. The network, by redesigning the feature aggregation in residual blocks and introducing the Split-Attention module, effectively captures inter-channel relationships, thereby enhancing feature extraction  capabilities.Mekruksavanich et al. \cite{mekruksavanich2023deep} propose a DL network with aggregation residual transformation called the ResNeXt model that can classify human activities based on inertial and stretch sensor data with satisfactory results. 

\subsection{Attention Mechanisms}
The introduction of attention mechanisms allows models to focus more on specific aspects, thereby improving performance. Essa et al. \cite{essa2023temporal} proposed the TCCSNet architecture, where the second branch consists of a set of convolutional and self-attention blocks to capture local and temporal features in sensor data. Pramanik et al. \cite{pramanik2023transformer}proposed an attention mechanism based on deep backward transformer to guide lateral residual features, ensuring that the model learns the optimal correlation information between spatial and temporal features. Liu et al. \cite{liu2023mag}combined a multi-scale residual network with gate mechanism and ECA attention mechanism, enhancing the capability of channel feature extraction to better differentiate various human movements in daily life.

\subsection{Data Augmentation}
Data augmentation plays a critical role in addressing data scarcity and enhancing model generalization in HAR. Several data augmentation techniques have been proposed for HAR. Cheng et al. \cite{cheng2023learning} employed contrastive supervision, using contrastive loss to supervise the intermediate layers of deep neural networks, proving effective in learning time-series invariance and improving classification accuracy. Wang et al. \cite{wang2023data} generated samples with varying distances, angles, and human motion velocities through operations such as distance shifting, angle rotation, and velocity simulation, thus enhancing the model's generalization across different scenarios. They further improved the accuracy of HAR tasks by combining contrastive learning with generative learning and employing automatic augmentation strategy search methods \cite{wang2023data}. Some data augmentation techniques, such as linear combination, scaling, and jittering, have preserved accurate labels in ConvLSTM networks, improving classification accuracy \cite{xu2023augmentation}. Diffusion models, a novel data augmentation method proposed in recent years, generate a wide range of synthetic sensor data that precisely represent the original by imposing conditional constraints on statistical properties \cite{shao2023study}. Today, diffusion models have been applied to data augmentation for imbalanced datasets, such as in epilepsy seizure prediction. To address imbalance issues, a new data augmentation method, DiffEEG, was introduced, utilizing diffusion models and demonstrating superiority over existing methods \cite{wang2023data1}. 
\section{Methodology}
This section introduces the proposed HAR framework. As shown in Figure \ref{fig1:ALL}, it consists of four parts: Data synthesis based on unsupervised diffusion model, Multi-brach network architecture of spatiotemporal attention interaction, Adaptive composite loss function and Embedded device deployment.

\subsection{Data Synthesis Based on Unsupervised Diffusion Model}
Wearable sensor data in HAR often faces the challenges of scarce labeled training data and annotation difficulties. These challenges can compromise the accuracy and robustness of HAR models in practical applications. Hence, addressing the scarcity and complexity of labeling training data is crucial for enhancing model performance. To overcome these issues, we generate diverse synthetic sensor data using the SF-DM model proposed by Si et al. \cite{zuo2023unsupervised}, which does not rely on labeled data for training, thereby improving the performance of HAR models.
\hangindent=0em
\par
\textbf{Method Implementation}: We construct an encoder-decoder framework: the encoder consists of three convolutional layers with 9×9 kernels and a max-pooling layer with a 2×2 kernel and stride of 2, to learn features from the input. The inputs to the convolutional layers are statistical features, noise data, and embedded diffusion steps. Before entering the convolutional layer, statistical features are projected to match the shape of the noise data. The output of the diffusion steps is added to the output of the noise data, then concatenated with the output of the statistical features processed by the convolutional layer, providing additional information. The decoder comprises an upsampling layer and a convolutional layer with a 9×9 kernel. The upsampling layer is responsible for restoring resolution before matching the previous layer, while the convolutional layer transfers information contained in one data point to multiple data points. Finally, an output projection layer matches the dimensions of the output from the diffusion model to the real input data. 
\hangindent=0em
\par
\textbf{Method Definitions}: Firstly, we extract statistical features from real data obtained from sensors. These features include the mean, standard deviation, Z-score $z=\frac{x-\mu}{\sigma}$ (where $\chi$ is the observed value, $\mu$ is the mean of all values, and $\sigma$ is the standard deviation of the sample), and skewness $\gamma=\mathbb{E}\left[\left(\frac{x-\mu}\sigma\right)^3\right]$. These features have the same length as the input sequence and are fully connected as $f$. Then, we input the sensor data to generate noisy data $\tilde{x}$. Next, we feed $\tilde{x}$ and the statistical features into a diffusion model, training the diffusion model to generate synthetic data by minimizing the reconstruction loss between the original real wearable sensor data and the generated data and pre-training the model. Subsequently, we calculate the statistical characteristics, fully connecting them to $f$, and train the diffusion model by minimizing the synthetic loss. Finally, the real sensor data is used to fine-tune the HAR classifier by minimizing the real loss, and the relevant data is output.
The mathematical formulations of the entire process are shown in equations (1):
\begin{equation}
\begin{aligned}\tilde{x}=x\times\sqrt{\beta[t]}&+\in x\sqrt{1-\beta[t]}\\
{L_{rec}(x, \tilde{x}, f;\theta_E)}&=\frac{\sum_{l=1}^n|D(x_l, f_l)-x_l|}n\\
{L_{syn}(\omega, f, y;\theta_C)}&=-\sum_{l\overset{c}{\operatorname*{=}}1}^{n_c}{y_llogC(E(\omega_{l^{}}, f_l))}\\
{L_{real}\left(x, y;\theta_C\right)}&=-\sum_{l=1}^ny_l\mathrm{log}C\left(x_l\right)\\
\end{aligned}
\end{equation}
In the above equations, $x$ represents the labeled real sensor data, $\beta$ represents the noise level, $t$ represents the diffusion step, and $\epsilon$ provides random noise with the same shape as the input real data. $\tilde{\chi}$ represents the input noisy data, $f$ represents the statistical features, $D$ represents the diffusion model with decoder and encoder, and $n$ is the number of training samples. $\gamma$ represents the corresponding class labels, $E$ represents the pretrained diffusion model, $C$ is the activity classifier, $\omega$ represents the random noise input to the diffusion model, and $n_\mathrm{c}$ represents the number of classes,$L_{rec}$ represents the reconstruction loss,$L_{syn}$ represents the synthetic loss,$L_{real}$ represents the real loss. 
\begin{figure*}
    \centering
    \includegraphics[width=1\linewidth]{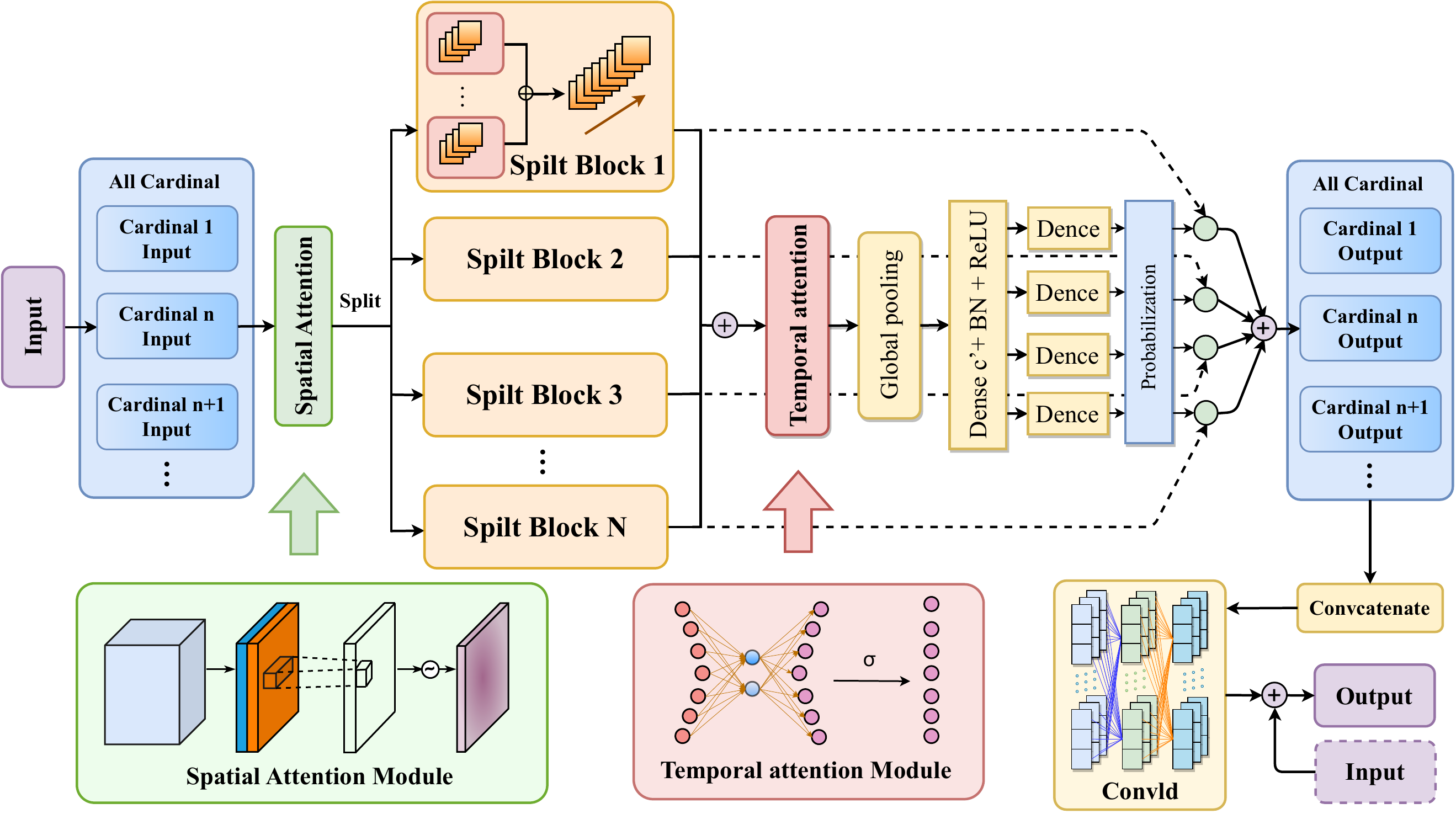}
    \caption{MSTI Model}
    \label{fig2:MSTI}
\end{figure*}
\subsection{Multi-Brach Network Architecture of Spatiotemporal Attention Interaction}
Models with split-attention can improve accuracy without increasing computational cost. Based on these advantages, our proposed MSTI model improves the split-attention branch number, the number of channels in each branch and intermediate layer of each residual block, and the network layer structure, aiming to improve computational efficiency. To better capture temporal and spatial features in the data, we introduce temporal attention and spatial attention on this basis. Thus, the designed MSTI model improves computational efficiency while extracting deep features from the data. Figure \ref{fig2:MSTI} shows the model structure of MSTI.
\hangindent=0em
\par
This is a Multi-Branch architecture network primarily comprised of Convolutional Neural Networks (CNNs). It is designed to focus on various aspects such as spatial and temporal features at different time periods and optimize feature maps through decomposition and reconstruction operations. In the preceding section, the generated data mentioned therein, along with the original data, are input into the MSTI (\textit{Multi-Stage Temporal Inception}) model. Features are divided into several groups, with the number of feature map groups determined by the cardinality hyperparameter, denoted as $K$. Another hyperparameter, the radix denoted as \(R\), represents the number of splits within a radix group. Therefore, the total number of feature groups is given by $G=KR$. We can apply a series of transformations \(\{F_1, F_2, ..., F_n\}\), where each group's intermediate representation is denoted as \(U_i = F_i(X)\), for \(i \in \{1, 2, ..., G\}\).

\textbf{Cardinality:}
Like ResNeSt, a combinatorial representation can be obtained by element-wise summation over multiple split radix combinations. The representation for $k$-th cardinal group is $\hat{U}^k=\sum_{j=R(k-1)+1}^{R k} U_j$, where $\hat{U}^k \in \mathbb{R}^{S  \times C / K}$ for $k \in 1,2, \ldots K$, and $S$ and $C$ are the block output featuremap sizes. Global contextual information with embedded channel-wise statistics can be gathered with global average pooling across spatial dimensions $D^k \in \mathbb{R}^{C / K}$. Here the $c$-th component is calculated as: $D_c^k=\frac{1}{S} \sum_{j=1}^S \hat{U}_c^k(i, j) .$

\textbf{Radix:}
A weighted fusion of the cardinal group representation $V^k \in \mathbb{R}^{S \times C / K}$ is aggregated using channel-wise soft attention, where each featuremap channel is produced using a weighted combination over splits. Then the $c$-th channel is calculated as: $V_c^k=\sum_{i=1}^Ra_i^k(c) U_{R(k-1)+i}$ , where $a_i^k(c)$ denotes a (soft) assignment weight given by:

$a_i^k(c)= \begin{cases}\frac{\exp \left(G_i^c\left(s^k\right)\right)}{\sum_{j=1}^R\left(\exp \left(G_j^c\left(s^k\right)\right)\right.} & \text { if } R>1, \\ \frac{1}{1+\exp \left(-G_i^c\left(s^k\right)\right)} & \text { if } R=1,\end{cases}$

\textbf{Feature Concatenation:}
The cardinal group representations are then concatenated along the channel dimension: \(V = \text{Concat}\{V_1, V_2, ..., V_K\}\). Similar to standard residual blocks, if the input and output feature maps have the same shape, the final output \(Y\) for each radix group is generated using shortcut connections.

\textbf{Method Process:}
The input feature map is partitioned into $KR$ groups, each with one $R$ index and one $K$ index. In this layout, groups with the same radix index are adjacent to each other. Following the partitioning of the $K$ index, as the radix grouping inherently possesses channel-wise attentive features, we introduce spatial attention mechanisms to complement it. We initiate our process by applying average pooling and max pooling operations along the temporal axis, concatenating them to forge an effective feature descriptor. Upon the concatenated features, a convolutional layer is applied to generate a spatial attention map, $M_s(F)\in \mathbb{R}^{S\times C}$, which encodes positions to be emphasized or suppressed. Channelinformation is aggregated through the use of two pooling operations, resulting in the formation of two feature maps: $F_{avg}^s \in \mathbb{R}^{S\times C}$ and $F_{max}^s \in \mathbb{R}^{S\times C}$ . Each represents the average pooling and max pooling features across channels, respectively. Subsequently, they are concatenated and convolved through a standard convolutional layer. 
\begin{equation}
\begin{aligned}&F=([F_{avg}^{s};F_{max}^{s}])\\
&M_{s}(F)=\sigma(f^{3\times1}\times F))
\end{aligned}
\end{equation}
Here, $\sigma$ represents the sigmoid function. We can then sum across different splits so that groups of feature maps with the same K index but different R indexes are fused together. Then we focus on temporal attention for the features. Here, our aim is to temporarily reduce the dependency between feature map groups with different \(R\) indices. Thus, we prioritize considering the information from each channel of the features. Each filter operates using a local receptive field, ensuring that after transformation, each unit cannot leverage context information from beyond that region. We compress the global spatial information into channel descriptors by generating channel-wise statistics. Formally, a statistic $Z \in \mathbb{R}^C$ is generated by shrinking the output $\mathbf{U}$ through its spatial dimensions $S$, such that the $c$-th element of $Z$ is calculated by: 
\begin{equation}
Z_c=\frac{1}{S} \sum_{i=1}^S u_c(i, j)
\end{equation}
Next, global pooling layer maintaining feature separation along the channel dimension. This is equivalent to globally pooling each individual radix group and then concatenating the results. Subsequently, two consecutive fully connected (FC) layers, each with a number of groups equal to the radix, are added after the pooling layer to predict the attention of each split module. The use of grouped FC layers ensures that each pair of FC layers is applied uniformly across each radix group. Through this implementation, the preceding 1 × 1 convolutional layers can be unified into one layer, and a single grouped convolution with an RK group number can be used to realize a 3 × 1 convolutional layer. Finally, the dispersed attention blocks are modularized using standard CNN operators and residual links are established with the original input features.
\subsection{Adaptive Composite Loss Function}
Class imbalance presents a significant challenge as it can adversely affect the classifier's training and generalization capabilities. To address this issue, this study introduces a composite loss function that integrates cross-entropy loss, focal loss, and label-smoothing regularization to enhance the model's adaptability and robustness to imbalanced data. The cross-entropy loss, a standard loss function for multi-class classification problems, optimizes the model by minimizing the disparity between the predicted probability distribution and the target distribution. The focal loss, by adjusting the parameters $\alpha$(sample weight)~and $\gamma $ (modulating contribution of easy-to-classify examples), intensifies the model's focus on minority classes. The introduction of label-smoothing regularization aims to mitigate the model's overconfidence in its predictions, thereby improving generalization performance. 

The model employs the Adam optimizer for parameter updates, which incorporates the mechanisms of first-order and second-order moment estimation with bias correction, as well as a weight decay strategy. The specific expressions are as follows:
\begin{equation}
\begin{aligned}
m_t&=\beta_lm_{t-1}+(1-\beta_l)g_t \\
v_t&=\beta_2v_{t-1}+(1-\beta_2)g_t^2 \\
\hat{m}_{t}&=\frac{m_{t}}{r-\beta_{t}^{t}}\\
\hat{v}_{t}&=\frac{v_{t}}{l-\beta_{2}^{t}}\\
\theta_t&=\theta_{t-1}\frac{\eta}{\sqrt{\overline{v}_t}+\epsilon}(\hat{m}_t+\lambda\theta_{t-1})
\end{aligned}
\end{equation}
 where $m_t$ and $v_t$ represent the estimates of the first and second moments, respectively; $\hat{m}_{t}$ and $\hat{v}_{t}$ are their bias-corrected values; $\theta_t$ is the parameter vector; $\eta$ is the learning rate; $\epsilon$ is a stabilizing term to prevent division by zero; and $\lambda$ is the weight decay coefficient. 

 At the end of each training batch, the composite loss is composed of the following parts:
\begin{equation}
\begin{aligned}
Loss_{ce}&=-\sum_{i=1}p\left(x_{i}\right)\mathrm{log}q\left(x_{i}\right)  \\
Loss_{fl}&=-\alpha_{t}(1-p_{t})^{\gamma}\log{(p_{t})} \\
Loss_{sl-nll}&=-\sum_{k=1}^K\log p(k)\left((1-\epsilon)\delta_{k, y}+\frac\epsilon K\right) \\
Loss_{total}&=\omega_{0}Loss_{sl-nll}+\omega_{1}Loss_{fl}+\omega_{2}Loss_{ce}
\end{aligned}
\end{equation}
Here, $\alpha_\mathrm{t}$ is a weight factor in the range $[0, 1];\gamma$ is a modulating factor; $e$ is a hyperparameten within $[0, 1];K$ is the number of label categories; $k$ represents a specific label; and $\omega _0$ , $\omega_1$, and $\omega_2$ are the weights of the loss functions. 

The strategy of integrating multiple different loss functions is now applied to various models. In order to better balance multiple loss functions, the most commonly used method is to assign different weights. This weight adjustment strategy is usually applied after the end of all model cycles. This adjustment is time-consuming and finding a suitable weight value is challenging. Therefore, we design a weight algorithm called adaptive algorithm, which is inspired by the concept of '' feedback '' in automatic control theory, that is, control according to the information of system output change. The algorithm is defined as automatically adjusting the weight value of the loss function based on the operation results of each cycle and using it for subsequent cycles. Specifically, if the accuracy of the model in the current cycle is higher than that in the previous cycle, the higher weight value will decrease slightly in the next cycle, while the lower weight value will increase slightly, and vice versa. The design of changing the weight value in real time may cause the fluctuation of the model accuracy, but the long training cycle will lead to the model falling into local optimum. These fluctuations can help the model jump out of local optimum. The weight update formula is as follows :
\begin{equation}
\begin{aligned}
\omega_0&=0. 5*(1-\omega_1)\\
\omega_1&=2-\tau-\frac1{acc+1e-8}\\
\omega_2&=0. 5*(1-\omega_1)\\
\end{aligned}
\end{equation}
where $\tau$ is the weight factor of [0, 1], and $acc$ is the current accuracy of the model. Ultimately, the trained model is evaluated using the validation set to calculate the model's
 classification accuracy and loss function value, which are used to adjust the model parameters. After model fine-tuning, the final model is assessed using the test set to determine the model's classification accuracy and the value of the composite loss function. 

\subsection{Embedded Device Deployment}
In recent years, many studies only use high-performance device testing methods, and when actually deploying HAR models, it is often necessary to consider various limitations of the deployed devices. In view of this, we built an embedded deployment system that is in line with the current level of wearable technology to test the practical feasibility of various methods. This idea was inspired by Zhang. Their deployment experiments were based on Raspberry Pi 3 and 4. Compare the inference delay time of the proposed module and the baseline model \cite{DanHAR,CE}. We hope to build a more comprehensive and detailed set of deployment experiments, We put the specific implementation details in the Actual deployment in the experimental part.

\section{Experimental Design}
The environment for model training and inference experiments in actual deployment experiments is as shown in the table\ref{tab:Experimental environment } below. \begin{table}[h] 
\centering
\renewcommand\arraystretch{1.2}
\tabcolsep=0.25cm
\caption{Experimental environment }
\label{tab:Experimental environment }

\begin{tabular}{>{\centering\arraybackslash}p{1.5cm}|>{\centering\arraybackslash}p{3cm}|>{\centering\arraybackslash}p{3cm}}
\toprule
\textbf{Experiment} & Base& Deploy\\
\cmidrule(r){1-3}
\textbf{PyTorch} & 2.0.0 & 2.2.2\\
\cmidrule(r){1-3}
\textbf{Python} & 3.8& 3.11.2\\
\cmidrule(r){1-3}
\textbf{Cuda} & 11.8 & N/A\\
\cmidrule(r){1-3}
\textbf{GPU} & 2 * RTX 4090 24G & N/A\\
\cmidrule(r){1-3}
\textbf{CPU} & 16 * Intel(R) Xeon(R) Gold 6430 & Quad-core 64-bit Arm Cortex-A76 CPU\\
\cmidrule(r){1-3}
\textbf{Memory} & 120GB & 8GB\\
\bottomrule

\end{tabular}\end{table}

\subsection{DataSets}
For the evaluation of our model's performance, we selected the WISDM and PAMAP2 datasets to facilitate an objective assessment of our method. Here are the specific details of the datasets:
\hangindent=0em
\par
\textbf{PAMAP2} \cite{reiss2012introducing}: This dataset for HAR was released in 2012 by the Reiss and Stricker research group. It comprises data collected from 9 participants performing 18 different physical activities. Participants wore Inertial Measurement Units (IMUs) on their wrists, chest, and ankles. These IMUs include a tri-axial accelerometer, gyroscope, and magnetometer, capturing various activities such as walking, cycling, and playing soccer. Each sample was manually labeled, noting the body posture and type of activity. 
\hangindent=0em
\par
\textbf{WISDM} \cite{kwapisz2011activity}: This is a dataset for HAR research released by the Database Systems Research Center at Pennsylvania State University in 2010. It contains sample data of 51 volunteers performing 6 common daily activities, including standing, sitting, walking, going up and down stairs, and lying down. The data were collected using tri-axial miniature accelerometers worn on the volunteers’ wrists. 
\hangindent=0em
\par
\textbf{OPPORTUNITY}\cite{R.Chavarriaga}: Realistic daily life activities of 12 subjects in a sensor-enriched environment were recorded. In total, 15 networked sensor systems, including 72 sensors in 10 modalities, are integrated on the environment and the body.

To match the datasets with the model test for evaluating the model's predictive outcomes, the data were preprocessed. The specific parameters of the datasets with respect to the MSTI-AD model are shown in Table \ref{tab:Dataset Processing Details }. 
\begin{table}
\centering
\renewcommand\arraystretch{1.2}
\tabcolsep=0.3cm
\caption{Dataset Details }
\label{tab:Dataset Processing Details }
\begin{tabular}{>{\centering\arraybackslash}p{2cm}|>{\centering\arraybackslash}p{1.5cm}>{\centering\arraybackslash}p{1.5cm}>{\centering\arraybackslash}p{1.5cm}}
\toprule
\textbf{Datasets} & PAMAP2 & WISDM  &OPPOTUNITY\\
\cmidrule(r){1-4}
\textbf{Sensor} & 3\&40& 3  &72 
\\
\cmidrule(r){1-4}
\textbf{Subject} & 9 & 29  &12 
\\
\cmidrule(r){1-4}
\textbf{Class} & 9\&12& 6  &18 
\\
\cmidrule(r){1-4}
\textbf{Window Size} & 171 & 90  &113 
\\\cmidrule(r){1-4}
 \cmidrule(r){1-4}\textbf{Batch Size}& 256& 512 &256
\\\cmidrule(r){1-4}
\textbf{Lr} & 0.001 & 0.005 &0.001 
\\\cmidrule(r){1-4}
 \textbf{Epoch} & 30& 50 &40\\
 \bottomrule
\end{tabular}\end{table}
\subsection{Evaluation Metrics}
To assess the performance of the proposed HAR model, we employed multiple evaluation metrics for a comprehensive assessment of our model. Accuracy, the core metric of interest, measures the proportion of correct classifications made by the model across all samples, providing a quick overview of overall performance. F1-weighted offers a synthesis of performance for multiclass classification. Lastly, we use G-mean to examine the model's recognition effectiveness on real data after training on samples that have undergone data augmentation, ensuring robust performance across all categories. The combined use of these metrics allows for a thorough understanding of the model's strengths and limitations in the task of HAR. 
\begin{equation}
\begin{aligned}
\text{Accuracy} &= \frac{\mathrm{TP} +\mathrm{TN}}{\mathrm{TP}+\mathrm{FN}+\mathrm{FP}+\mathrm{TN}}\\
\text{Precision} &= \frac{\mathrm{TP}}{\mathrm{TP}+\mathrm{FP}}\\
\text{Recall} &= \frac{\mathrm{TP}}{\mathrm{TP}+\mathrm{FN}}\\
\text{F1-macro} &= \frac{2 \times (\text{Precision} \times \text{Recall})}{\text{Precision} + \text{Recall}}\\
\text{F1-weighted} &= \sum_{i} \frac{2 \times \omega_i \times (\text{Precision}_i \times \text{Recall}_i)}{\text{Precision}_i + \text{Recall}_i}\\
\text{G-mean}&=\sqrt{\frac{\mathrm{TP}}{\mathrm{TP}+\mathrm{FN}}+\frac{\mathrm{TN}}{\mathrm{TN}+\mathrm{FP}}}
\end{aligned}
\end{equation}
Where TP and TN  represent the number of true positives and true negatives, respectively, while FN and FP are the numbers of false negatives and false positives. Precision represents the average precision across all labels, and Recall represents the average recall rate across all labels.  $\omega_{i}$ is the proportion of the $i$ class samples. 
\section{Results \& Discussion}
\subsection{Ablation Study}
We chose to conduct ablation experiments on the WISDM data set. Compared with other complex data sets with high user distribution differences, the ablation experiment results are relatively stable and can reduce the workload to a certain extent. We disassembled MSTI to test the effectiveness of each module. ''Base'' represents the original model without adding additional attention mechanisms, and only contains the multi-branch structure of cardinality and radix grouping. Then we add spatial and temporal attention modules respectively, and compare them with the spatio-temporal interaction module. ''Composite loss '' refers to the use of unweighted composite function combinations to optimize model training. In view of the fact that there are few studies on these three loss functions, we selected two weight combinations that perform well in the model for experiments.In combination 1, $\omega_{0}$, $\omega_{1}$ and $\omega_{2}$ were 0.3, 0.4 and 0.3, respectively. In combination 2,  $\omega_{0}$, $\omega_{1}$ and $\omega_{2}$were 0.15, 0.7 and 0.15, respectively. ''AD '' refers to the adaptive algorithm  for weighted training optimization, which is a synonym for MSTI-AD. The purpose of this practice is to verify the effectiveness of the weight algorithm.

In fact, adding a spatial or temporal attention module alone cannot achieve significant gains in the model. We believe this may be due to over-emphasis on a single dimension, resulting in information loss or additional noise in the process, resulting in the model being unable to be correct. Understand the relationship between spatiotemporal characteristics. In contrast, MSTI has the interaction of temporal and spatial attention, resulting in substantial performance improvements. In addition, there is no difference in accuracy and G-mean using the composite loss function with or without weights.For unweighted, combination 1 and combination 2, the decrease of F1 index shows that the recognition rate of the model for different categories is significantly different.AD solves this problem well by dynamically adjusting the weight of each loss function. Finally, Diffusion Data brings some gains to the final results, especially in the G-mean indicator, which can show that the generated data is relatively reasonable in the actual physical space. However, the performance gain is not obvious on WISDM. We have conducted extensive and rigorous comparisons in subsequent related work comparisons, and we can clearly see the gain of Diffusion Data on the model.
\begin{table}
    \centering
    \renewcommand\arraystretch{1.2}
\tabcolsep=0.3cm
        \caption{Ablation Experiment}
    \label{tab:Ablation Experiment}
    \begin{tabular}{ccccc} 
    \toprule
         \multicolumn{2}{c}{\textbf{Model}}&  \multicolumn{3}{c}{\textbf{WISDM}}\\ 
         \cmidrule(r){1-5}Main&   Additional&  ADC&  F1-w& G-mean\\

         \cmidrule(r){1-5}\multirow{4}{*}{Base}&  

{$/$}&  97.67&  97.67& 98.08
\\ 
         \cmidrule(r){2-5}&

Spatial Attention&  97.73&  97.72& 97.98
\\ 
         \cmidrule(r){2-5}&

Temporal Attention&  97.6&  97.62& 98.01
\\ 
         
\cmidrule(r){1-5}\multirow{4}{*}{MSTI}&  
/&  98.58&  98.58& 98.98
\\ 
         
\cmidrule(r){2-5}&  Compound Loss&  98.45&  97.31& 98.25
\\
 \cmidrule(r){2-5}& Combinations 1& 98.47& 97.91&98.45\\
 \cmidrule(r){2-5}& Combinations 2& 98.53& 98.09&98.6\\ 
         \cmidrule(r){2-5}&  AD&  98.64&  98.56& 98.88
\\ 
         \cmidrule(r){1-5}{MSTI-AD}&  Diffusion Data
&  98.84&  98.79& 99.02
\\ \bottomrule
    \end{tabular}
    
\end{table}
\begin{table*}
\centering
\renewcommand\arraystretch{1.2}
\tabcolsep=0.3cm

\caption{Comparison of related work}
\label{tab:Comparison}
\begin{tabular}{c|cccccc}     
\toprule
\multirow{2}{*}{\textbf{Model}}& \multicolumn{2}{c}{\textbf{PAMAP2(100\%)}} & \multicolumn{2}{c}{\textbf{PAMAP2(50\%)}} & \multicolumn{2}{c}{\textbf{WISDM}} \\ 
\cmidrule(r){2-3}\cmidrule(r){4-5}\cmidrule(r){6-7} & Accuracy & F1-weighted & Accuracy & F1-weighted & Accuracy & F1-weighted \\ 
\cmidrule(r){1-1}\cmidrule(r){2-3}\cmidrule(r){4-5}\cmidrule(r){6-7}CNN\cite{zeng2014convolutional}& 64.19& 60.63& 60.29& 58.25& 93.31 & 93.51 \\
\cmidrule(r){1-1}\cmidrule(r){2-3}\cmidrule(r){4-5}\cmidrule(r){6-7}LSTM\cite{xia2020lstm}& 60.86& 57.40& 55.10& 47.37& 96.71 & 96.68 \\
\cmidrule(r){1-1}\cmidrule(r){2-3}\cmidrule(r){4-5}\cmidrule(r){6-7}LSTM-CNN\cite{xia2020lstm}& 57.71& 53.31& 52.57& 48.68& 95.90& 95.97 \\  
\cmidrule(r){1-1}\cmidrule(r){2-3}\cmidrule(r){4-5}\cmidrule(r){6-7}CNN-GRU\cite{dua2021multi}& 61.43& 54.92& 60.48& 52.15& 94.95 & 96.21 \\
\cmidrule(r){1-1}\cmidrule(r){2-3}\cmidrule(r){4-5}\cmidrule(r){6-7}SE-Res2Net\cite{gao2019res2net}& 70.29& 67.62& 66.37& 62.46& 95.52 & 95.56 \\
\cmidrule(r){1-1}\cmidrule(r){2-3}\cmidrule(r){4-5}\cmidrule(r){6-7}ResNeXt\cite{mekruksavanich2022deep}& 62.48& 60.02& 58.57& 53.77& 96.67 & 96.66 \\  
\cmidrule(r){1-1}\cmidrule(r){2-3}\cmidrule(r){4-5}\cmidrule(r){6-7}Gated-Res2Net\cite{yang2020gated}& 70.48& 68.75& 65.52& 64.31& 97.02 & 97.02 \\
\cmidrule(r){1-1}\cmidrule(r){2-3}\cmidrule(r){4-5}\cmidrule(r){6-7}Rev-Attention\cite{pramanik2023transformer}& 63.90& 61.59 & 58.86& 57.37& 97.46 & 97.49 \\ 
\cmidrule(r){1-1}\cmidrule(r){2-3}\cmidrule(r){4-5}\cmidrule(r){6-7}MAG-Res2Net\cite{liu2023mag}& 72.24& 70.72& 67.34& 65.87& 98.32 & 98.42 \\
 \cmidrule(r){1-1}\cmidrule(r){2-3}\cmidrule(r){4-5}\cmidrule(r){6-7}HAR-CE\cite{CE}& 65.90& 58.55& 61.81 & 58.35 & 97.72 &97.7 
\\
 \cmidrule(r){1-1}\cmidrule(r){2-3}\cmidrule(r){4-5}\cmidrule(r){6-7}ELK\cite{ELK}& 72.95 & 72.77 & \textbf{72.57}& \textbf{72.42}& 98.05&98.06 
\\
 \cmidrule(r){1-1}\cmidrule(r){2-3}\cmidrule(r){4-5}\cmidrule(r){6-7}DanHAR\cite{DanHAR}& 71.52 & 70.27& 70.57& 68.18 & 95.57 &95.64 \\ 
\cmidrule(r){1-1}\cmidrule(r){2-3}\cmidrule(r){4-5}\cmidrule(r){6-7}\textbf{MSTI-AD}& \textbf{73.05}& \textbf{72.80}& 69.29& 68.92& \textbf{98.84}& \textbf{98.79}\\
\bottomrule
\end{tabular}
\end{table*}
\begin{figure}
    \centering
    \includegraphics[width=1\linewidth]{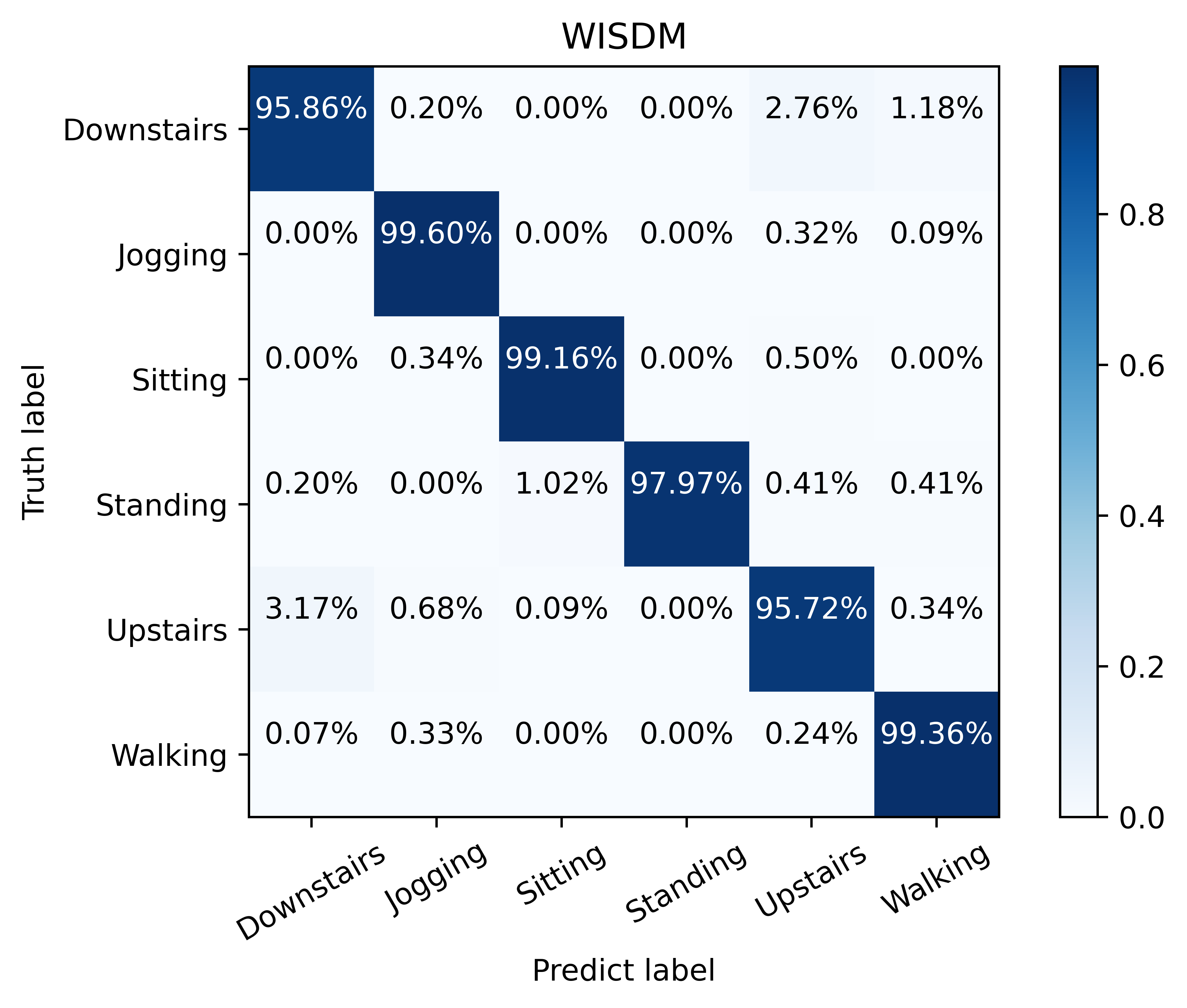}
    \caption{Confusion matrices on the WISDM}
    \label{fig:ConfusionMatrix}
\end{figure}
\begin{figure}[h]
    \centering
    \includegraphics[width=0.8\linewidth]{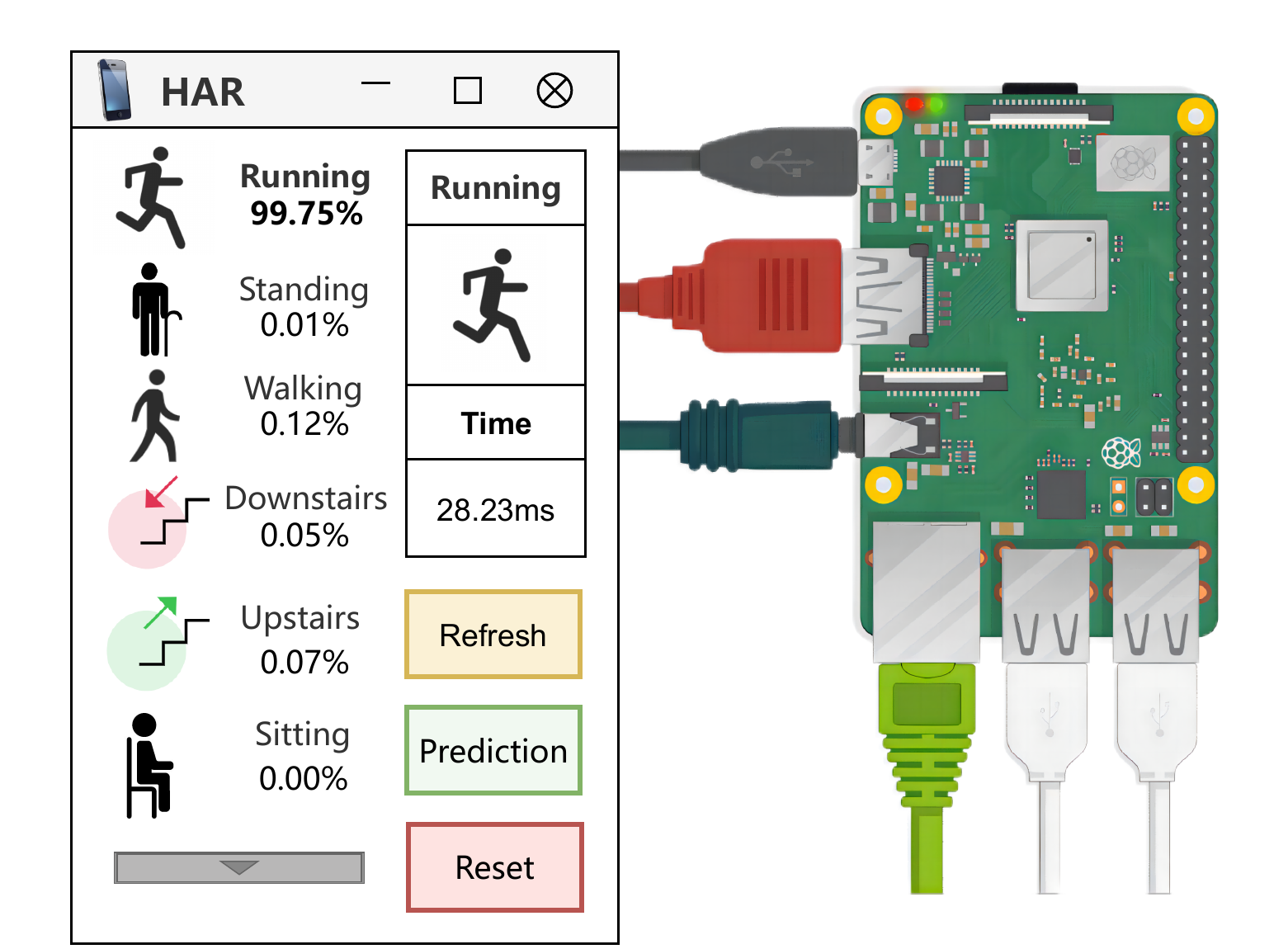}
    \caption{Model deployment.}
    \label{fig:Model deployment}
\end{figure}
\begin{table*}
\centering
\renewcommand\arraystretch{1.2}
\tabcolsep=0.25cm
\caption{Classification results of the WISDM  dataset. }
\label{tab:WISDM Test}
\begin{tabular}{>{\centering\arraybackslash}p{1.9cm}|llllcccccc}   
\toprule
\multirow{2}{*}{\textbf{WISDM}}&   \multicolumn{2}{c}{\textbf{LSTM-CNN}}&\multicolumn{2}{c}{\textbf{CNN-GRU}}&\multicolumn{2}{c}{\textbf{SE-Res2Net}} & \multicolumn{2}{c}{\textbf{ResNeXt}} & \multicolumn{2}{c}{\textbf{Gated-Res2Net}}   \\
 \cmidrule(r){2-3}\cmidrule(r){4-5} \cmidrule(r){6-7}\cmidrule(r){8-9}\cmidrule(r){10-11}
 &  ADC& F1-m&ADC& F1-m&ADC& F1-m& ADC& F1-m& ADC& F1-m\\
\cmidrule(r){1-11}
Downstairs &   89.29& 90.99&90.37& 93.93
&80.16  & 88.56  & 90.43& 86.77  & 87.14  & 89.98  \\
\cmidrule(r){1-1}\cmidrule(r){2-3}\cmidrule(r){4-5} \cmidrule(r){6-7}\cmidrule(r){8-9}\cmidrule(r){10-11}
Jogging &   98.96& 99.25&99.19& 99.31
&99.28  & 99.30& 99.08  & 99.02  & 98.86  & 99.17  \\
\cmidrule(r){1-1}\cmidrule(r){2-3}\cmidrule(r){4-5} \cmidrule(r){6-7}\cmidrule(r){8-9}\cmidrule(r){10-11}
Sitting &   98.31& 98.98&98.30& 99.49
&98.31  & 98.32  & 99.32  & 98.82  & \textbf{99.66}& 98.39  \\
\cmidrule(r){1-1}\cmidrule(r){2-3}\cmidrule(r){4-5} \cmidrule(r){6-7}\cmidrule(r){8-9}\cmidrule(r){10-11}
Standing &   98.15& 97.15&98.56& 97.56
&98.13  & 98.34  & 97.95  & 98.45  & 97.56  & 98.56  \\
\cmidrule(r){1-1}\cmidrule(r){2-3}\cmidrule(r){4-5} \cmidrule(r){6-7}\cmidrule(r){8-9}\cmidrule(r){10-11}
Upstairs &   90.76& 91.07&91.36& 89.46
&96.43& 89.13  & 80.44  & 87.71  & 86.42  & 91.00  \\
\cmidrule(r){1-1}\cmidrule(r){2-3}\cmidrule(r){4-5} \cmidrule(r){6-7}\cmidrule(r){8-9}\cmidrule(r){10-11}
Walking &   98.49& 97.77&98.46& 98.06
&98.81& 98.92  & 99.52  & 98.62  & 99.27  & 99.16  \\
%\bottomrule
\end{tabular}%\end{table*}

%\begin{table*}
%\centering
%%\caption{Classification results of the WISDM  dataset(2). }
%\label{tab:WISDM Test(2)}

\begin{tabular}{ccccccllllll}   
\toprule
      \multicolumn{2}{c}{\textbf{Rev-Attention}} &  \multicolumn{2}{c}{\textbf{MAG-Res2Net}}& \multicolumn{2}{c}{\textbf{HAR-CE}
} &  \multicolumn{2}{c}{\textbf{ELK}}& \multicolumn{2}{c}{\textbf{DanHAR}}& \multicolumn{2}{c}{\textbf{MSTI-AD}}\\
\cmidrule(r){1-2}\cmidrule(r){3-4} \cmidrule(r){5-6}\cmidrule(r){7-8}\cmidrule(r){9-10}\cmidrule(r){11-12}
           ADC& F1-m& ADC& F1-m& ADC& F1-m & ADC& F1-m & ADC& F1-m & ADC&F1-m 
\\
\cmidrule(r){1-12}
           88.54  & 92.46& 91.25& 91.46& 91.46 & 93.28 & 91.85 & 93.19 & 92.60 & 85.99 
& \textbf{95.86}&\textbf{95.91} 
\\
\cmidrule(r){1-2}\cmidrule(r){3-4} \cmidrule(r){5-6}\cmidrule(r){7-8}\cmidrule(r){9-10}\cmidrule(r){11-12}
           99.20  & 99.14   & 99.20& 98.87& 99.42 & 99.32 & 99.51 & 99.19 & 97.85 & 98.65 
& \textbf{99.60}&\textbf{99.42} 
\\
\cmidrule(r){1-2}\cmidrule(r){3-4} \cmidrule(r){5-6}\cmidrule(r){7-8}\cmidrule(r){9-10}\cmidrule(r){11-12}
           99.15  & 99.07& 98.33& 98.56& 99.49 & 98.73 & 99.66 & 99.16 & 99.83 & 99.07 
& 99.16&\textbf{99.08} 
\\
\cmidrule(r){1-2}\cmidrule(r){3-4} \cmidrule(r){5-6}\cmidrule(r){7-8}\cmidrule(r){9-10}\cmidrule(r){11-12}
           98.34  & 98.47 & \textbf{98.74}& 98.27& 97.36 & 98.46 & 97.96 & 98.77 & 98.36 & 99.07 
& 97.97&\textbf{98.97} 
\\
\cmidrule(r){1-2}\cmidrule(r){3-4} \cmidrule(r){5-6}\cmidrule(r){7-8}\cmidrule(r){9-10}\cmidrule(r){11-12}
           95.44  & 92.36   & 93.46& 91.23& \textbf{97.09} & 92.50 & 94.57 & 94.04 & 77.23 & 85.35 
& 95.72&\textbf{95.56} 
\\
\cmidrule(r){1-2}\cmidrule(r){3-4} \cmidrule(r){5-6}\cmidrule(r){7-8}\cmidrule(r){9-10}\cmidrule(r){11-12}\cmidrule(r){11-12}
           99.34  & 99.33   & 98.71& 98.64& 97.84 & 98.64 & 99.13 & 99.16 & \textbf{99.83 }& 97.46 & \textbf{99.36}&99.43 \\
\bottomrule
\end{tabular}
\end{table*}

\subsection{Comparison of Related Work}
We compared our models against a range of methods, including traditional CNN and LSTM network frameworks, as well as models with state-of-the-art performance models. Furthermore, to accurately describe the performance of our method, networks related to our strategy such as SE-Res2Net, Gated-Res2Net, MAGRes2Net were also considered for comparison. We used the WISDM and PAMAP2 data sets for evaluation. Considering the requirements for data enhancement, we adopted a method inspired by \cite{zuo2023unsupervised} and selected the three-axis acceleration data from the PAMAP2 data set for enhancement and testing. The test results are shown in table \ref{tab:Comparison}. 

Notably, MSTI-AD demonstrates higher accuracy on the PAMAP2 dataset with 100\% data volume. On PAMAP2 with 50\% of the data volume, the comprehensive performance of MSTI-AD has also reached an extremely high level, surpassing all models except ELK. MSTI-AD achieves an accuracy of 98. 84\% on WISDM, exceeding the highest benchmark previously set by other models. These results verify the superior performance of our model on both simple and complex datasets. 

The confusion matrix for our model on the WISDM dataset is depicted in Figure \ref{fig:ConfusionMatrix}. MSTI-AD achieves higher accuracy across a broader spectrum of categories, effectively mitigating issues stemming from category confusion. Additionally, in Table \ref{tab:WISDM Test}, we showcase the recognition capabilities of existing high-performance models for fine-grained categories, including both accuracy and F1 scores.

\subsection{Actual Deployment}

We designed four complexity index test experiments on the WISDM and OPPO data sets for the baseline and SOTA models, that is, the model is in the process of inferring each data set. model parameter size, memory consumption, inference latency, and inference latency of a single fragment. Since PyTorch can currently support the deployment of deep learning models on Raspberry Pi 5, this work was performed on Raspberry Pi 5 equipped with a quad-core Cortex-A76  (ARM) 64-bit SoC @ 2.4 GHz and 8gb LPDDR4X-4267 SDRAM, and the pytorch version is 2.2.2. Figure \ref{fig:Model deployment} shows our software interfaceFor reference, here are some examples of smartphone chips with comparable overall performance to this processor: Snapdragon 865/Apple A12/Exynos 1080/Dimensity 1200 or Kirin 9000SL. 
\begin{figure}
    \centering
    \includegraphics[width=1.05\linewidth]{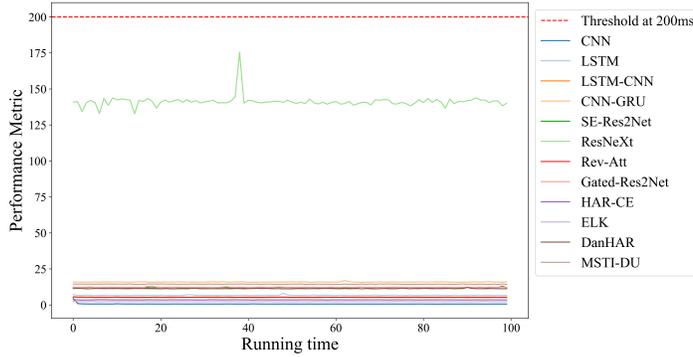}
    \caption{Inference delay of each model on the WISDM dataset}
    \label{fig:inference delay:wisdm}
\end{figure}

\begin{figure}[h]
    \centering
    \includegraphics[width=1.05\linewidth]{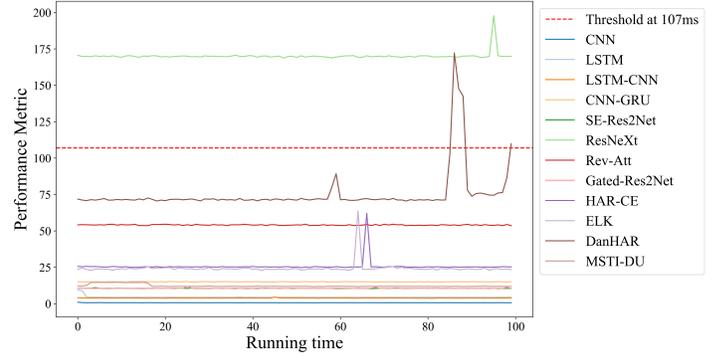}
    \caption{Inference delay of each model on the PAMAP2 dataset}
    \label{fig:inference delay:oppo}
\end{figure}
\begin{table}[h]
\centering

\tabcolsep=0.4cm
\caption{Complexity analysis. }
\label{tab:Complexity analysis}

\begin{tabular}{c|>{\centering\arraybackslash}p{0.8cm}>{\centering\arraybackslash}p{0.9cm}>{\centering\arraybackslash}p{0.8cm}>{\centering\arraybackslash}p{0.9cm}}
\toprule
\multirow{2}{*}{\textbf{Model}}&\multicolumn{2}{c}{\textbf{WISDM}}&\multicolumn{2}{c}{\textbf{OPPORTUNITY}}\\
\cmidrule(r){2-3}\cmidrule(r){4-5}& Memory& Param. & Memory& Param. \\
\cmidrule(r){1-1}\cmidrule(r){2-3}\cmidrule(r){4-5}
CNN & 766.23
& 1.04E+05 & 744.52& 1.21E+05 \\
\cmidrule(r){1-1}\cmidrule(r){2-3}\cmidrule(r){4-5}
LSTM & 759.80
& 1.65E+05 & 748.44& 3.85E+05 \\
\cmidrule(r){1-1}\cmidrule(r){2-3}\cmidrule(r){4-5}
CNN-GRU & 787.05
& 4.38E+06 & 773.17& 4.39E+06 \\
\cmidrule(r){1-1}\cmidrule(r){2-3}\cmidrule(r){4-5}
LSTM-CNN & 823.00
& 2.12E+06 & 788.69& 2.13E+06 \\
\cmidrule(r){1-1}\cmidrule(r){2-3}\cmidrule(r){4-5}
SE-Res2Net & 787.67
& 1.60E+06 & 772.41& 1.61E+06 \\
\cmidrule(r){1-1}\cmidrule(r){2-3}\cmidrule(r){4-5}
ResNeXt & 889.78
& 2.20E+07 & 849.25& 2.21E+07 \\
\cmidrule(r){1-1}\cmidrule(r){2-3}\cmidrule(r){4-5}
Rev-Att & 862.03
& 4.33E+05 & 801.97& 3.80E+05 \\
\cmidrule(r){1-1}\cmidrule(r){2-3}\cmidrule(r){4-5}
Gated-Res2Net & 773.84
& 1.60E+06 & 769.53& 1.61E+06 \\
\cmidrule(r){1-1}\cmidrule(r){2-3}\cmidrule(r){4-5}
HAR-CE& 788.73
& 4.17E+05 & 779.84
& 3.49E+06\\
\cmidrule(r){1-1}\cmidrule(r){2-3}\cmidrule(r){4-5}
ELK& 782.70
& 2.41E+05& 751.84
&7.57E+05\\
\cmidrule(r){1-1}\cmidrule(r){2-3}\cmidrule(r){4-5}
DanHAR& 790.66
& 2.35E+06& 814.09
&4.46E+06
\\
\cmidrule(r){1-1}\cmidrule(r){2-3}\cmidrule(r){4-5}
MSTI-AD&  718.50&   3.47E+05
&   751.30&  1.09E+06
\\
\bottomrule

\end{tabular}

\end{table}
In order to ensure the stability of the test, we continuously tested all models 100 times. The evaluated indicators include running time, memory usage. We did not record the results of running these models on the entire data set, because most of them were unable to complete this test, and we believe that such a test is of no practical value without adapting to the memory management method. Table \ref{tab:Complexity analysis} shows the average results of ten rounds of 100 tests on WISDM and OPPO datasets. It can be clearly seen that the memory consumption of our proposed method is lower than most state-of-the-art work, while the model parameters are fewer among them.

In order to further explore the absolute efficiency of the model, we set up an inference delay test to test the time it takes for the model to infer a single sample and continuously infer 100 samples. We referred to Zhang's inference delay test method and tested the sensor data through a sliding step window \cite{NNU2023}. They selected a 10s action clip , the step size of the sliding window is 95\% of the total length of the window, which is 500ms, and it is considered that less than 500ms meets the practical standard. However, after our testing, we found that due to the improvement of the performance of the deployment equipment, most models can easily achieve this standard. However, for the sliding window applicable to the public data set, compared with the commonly used window length of the data set, the previously selected 10s is a very huge value. For example, the commonly used segment length in WISDM is 4s, which is the most commonly used sliding window length in general. Then the corresponding window step size is 200ms, which means that the system will segment the data and predict every 200 milliseconds. New sample. We apply this approach to each dataset, i.e. the segments are segmented exactly as in the base experiment, and the latency requirement for model inference for a single segment is set to a 5\% window of the length of the individual segment to ensure the model is segmented in the next segment. 

The experimental results are shown in Figure  \ref{fig:inference delay:wisdm} and Figure  \ref{fig:inference delay:oppo}. On the WISDM data set, all the models we tested can complete the task within the limited time, and most of the models are concentrated in the lower time interval. This is consistent with our expectations for the performance of state-of-the-art embedded devices and the complexity of data sets. On the OPPORTUNITY data set, there is a clear gap in the efficiency of each model. The basic model still has a lower time, and some lightweight models, such as HAR-CE, ELK and our MSTI-AD can also be maintained at Within 25ms, while ResNeXt completely exceeded the time limit.

\section{Conclusion} \label{Conclusion}
In this study, we proposed a set of efficient HAR process methods, which effectively solved the pain points of HAR in data, deep feature extraction and actual deployment. The results obtained by this method through ablation experiments have a certain interpretability, and the advanced performance of the method is proved through comparative experiments. It is worth mentioning that we have built an embedded test system suitable for the performance of current wearable devices and extensively verified the rationality of the HAR model, making the HAR experiment not only simulated on high-performance devices, but also more To be realistic and unprecedented, we will make the code of the proposed method and the embedded test system publicly available.

\section*{Acknowledgment}\label{Acknowledgment}
This work was supported by National Natural Science Foundation of China (62072089); Fundamental Research Funds for the Central Universities of China (N2116016, N2104001 and N2019007).

% References
\bibliographystyle{IEEEtran}

\bibliography{jsen} %IEEEabrv instead of IEEEfull
%\vspace{-1cm}

\end{document}